\documentclass{aastex62}
\usepackage{CJK}

\usepackage{bm}
\usepackage{amsmath}

\received{}
\revised{}
\accepted{}
\submitjournal{ApJ}

\shorttitle{tearing instability in 2D current sheet}
\shortauthors{Shi et al.}
\begin{document}
\begin{CJK*}{UTF8}{gbsn}

\title{Marginal Stability of Sweet-Parker Type Current Sheets at Low Lundquist Numbers}

\correspondingauthor{Chen Shi}
\email{cshi1993@g.ucla.edu}

\author[0000-0002-2582-7085]{Chen Shi (时辰)}  %
\affiliation{EPSS, UCLA, Los Angeles, CA 90095, USA}

\author{Marco Velli}
\affiliation{EPSS, UCLA, Los Angeles, CA 90095, USA}
%\collaboration{(AAS Journals Data Scientists collaboration)}

\author{Anna Tenerani}
\affiliation{EPSS, UCLA, Los Angeles, CA 90095, USA}



\begin{abstract}
Magnetohydrodynamic simulations have shown that a non-unique critical Lundquist number $S_c$ exists, hovering around $S_c \sim 10^4$, above which threshold Sweet-Parker type stationary reconnecting configurations become unstable to a fast tearing mode dominated by plasmoid generation. It is known that the flow along the sheet plays a stabilizing role, though a satisfactory explanation of the non-universality and variable critical Lundquist numbers observed is still lacking. Here we discuss this question using 2D linear MHD simulations and linear stability analyses of Sweet-Parker type current sheets in the presence of background stationary inflows and outflows at low Lundquist numbers ($S\le 10^4$).  Simulations show that the inhomogeneous outflow stabilizes the current sheet by stretching the growing magnetic islands and at the same time evacuating the magnetic islands out of the current sheet. This limits the time during which fluctuations which begin at any given wave-length can remain unstable, rendering the instability non-exponential. We find that the linear theory based on the expanding-wavelength assumption works well for $S$ larger than $\sim 1000$. However we also find that the inflow and location of the initial perturbation also affect the stability threshold.
%We estimate that the critical $S$ for a strongly perturbed current sheet is about several thousands, depending on the pattern of the background magnetic field but we do not have a universal criterion for determining the critical $S$.
\end{abstract}

\keywords{magnetic reconnection, magnetohydrodynamics}



\section{Introduction} \label{sec_intro}
Magnetic reconnection is a phenomenon where magnetic energy is converted into the thermal and kinetic energy of the plasma via an induction electric field, leading to a reconfigured  magnetic field topology. The classic stationary reconnection model was developed by Sweet and Parker (SP) \citep{parker1957, sweet1958}, who studied magnetic field annihilation in a two-dimensional current layer with a macroscopic half-length $L$. It predicts a reconnection rate $R \sim S^{-1/2}$ where $S$ is the Lundquist number defined as $S=L V_{Au}/\eta$ ($V_{Au}$ and $\eta$ are the upstream Alfv\'en speed and the magnetic diffusivity). For current sheets in space plasmas, $S$ is usually very large (e.g. $\sim 10^{13}$ in the solar corona), and the typical time-scales are much longer than the duration of explosive events such as flares: thus, even though the inverse aspect ratio of the current sheet is small, $a/L \simeq S^{-1/2}$ ($a$ is the half-thickness of the current sheet), the SP model cannot explain explosive reconnection properly.

However, simulations by \citep{biskamp1986} showed that the SP stationary reconnecting current sheet becomes unstable to an extremely fast super-tearing, or plasmoid instability, once a critical value for the Lundquist number $10^4\le S_c \le10^5$ is exceeded. \citet{loureiro2007} pointed out, (after \citep{tajima2002}), that the maximum growth rate of the instability scales as $\sim S^{1/4}$ . The positive scaling of the instability growth rate with Lundquist number, and its divergence in the ideal limit, led
\citet{puccivelli2014} to point out that the only conclusion of such stability studies should be that large-$S$ SP current sheets should not form in nature in the first place. They identified a critical aspect ratio scaling $L/a\sim S^{1/3}$ above which current sheets become unstable even in the limit of ideal MHD. The critical aspect ratio scaling separates current sheets for which the tearing instability diverges, from those in which the growth rate of the tearing mode goes to zero as $S\rightarrow \infty$. At the critical scaling the tearing instability becomes ideal -- the growth rate becomes independent of the Lundquist number. Resistive MHD simulations by \citet{tenerani2015} and \citet{landi2015} have confirmed the existence and relevance of the critical aspect ratio even for nonlinear simulations.

The works mentioned above all focus on large Lundquist numbers ($S \ge 10^5$), though as mentioned above the early MHD simulations suggested that at $S\le 10^4$ the current sheet is actually stable and laminar \citep{biskamp1986}. \citet{bulanov1978} proposed that, unlike static current sheets which are always unstable (albeit to the very slow, for macroscopic current layers, tearing instability), the inhomogeneous outflow of the SP configuration can have a stabilizing effect leading to the Lundquist number threshold for the current sheet stability. \citet{biskamp2005} further analyzed Bulanov's model and estimated roughly the threshold for the aspect ratio $L/a$ to be $\sim 10^2$, corresponding to a Lundquist number $S \sim 10^4$. However, \citet{loureiro2007} claimed that the outflow only plays a mildly stabilizing role. Indeed it was further suggested (\citet{loureiro2013}) that stabilization might be provided by a more mathematical criterion, namely that in order for the tearing instability to develop in a SP current sheet, the inner singular layer which tearing develops should be significantly narrower, i.e. by about a factor of $3$, than the overall current sheet thickness. Remark that this criterion might be important in invalidating the classical asymptotic expansions \citep{furth1963} with which the analytical approximation of the dispersion relation for resistive modes was originally derived, but arguably has no real physical meaning: indeed, as we will discuss further in the conclusions, it can not justify the observed SP stabilization.

More recently the 2D linear MHD simulation by \citet{nietal2010} confirmed that, though the effect of flows is negligible when $S$ is large, its importance increases with decreasing $S$. In their simulations, initial random perturbations grow linearly at first but then saturate. They determine a critical Lundquist number $S \sim 2000$ at which the random perturbation only grows by a factor of 5. The simulation by \citet{nietal2010} is periodic in the outflow direction but employs a ``viscous buffer region'' at the outer boundaries where viscosity is very large to damp out perturbations: the need to have vanishing boundary conditions and associated damping regions might therefore influence the stability criterion significantly.
Indeed, realistic background configurations are periodic neither in the outflow nor inflow directions, rather the flow is accelerated into a jet diverging from the sheet at the background Alfv\'en speed.

In this study, we carry out linear MHD simulations and linear stability analysis of 2D Sweet-Parker type current sheets at low Lundquist numbers ($S \le 10^4$). We use open boundary conditions in both the inflow and outflow directions in the simulations. Our result shows that the inhomogeneous outflow stretches the magnetic islands forming and growing in the current sheet while at the same time evacuating them from the simulation domain. Both stretching and evacuation play a stabilizing role. In addition, the stretching effect changes the linear instability from the exponential behavior of the static instability, to one in which growth is only exponential over a limited time-interval. We confirm that the overall growth of fluctuations is limited, the reason being that the effective wave-number of unstable perturbations decreases over time limiting the overall period of instability.

The following section describes our numerical method; Section (\ref{sec:num_result_dawson}) describes the numerical results. Sections (\ref{sec:linear_analysis}) and (\ref{sec:initial_pert}) then describe SP stability while Section (\ref{sec:num_result_tanh}) shows numerical results for a different background configuration. Section (\ref{sec:conclusion}) describes the relevant outlook of our results.

\section{Numerical Method and Simulation Setup}\label{sec:num_setup}
The governing equations for the resistive stability of SP current sheets are given by the linearized resistive MHD equations (and the adiabatic equation for closure - this decouples the dynamic instability from potential thermodynamic effects):
\begin{subequations}\label{linear_MHD_eq_set}
  \begin{equation}\label{eq:linear_MHD_eq_set:rho1}
    \frac{\partial \rho_1}{\partial t} + \mathbf{u_0} \cdot \nabla \rho_1 + \rho_0 \nabla \cdot \mathbf{u_1} + \mathbf{u_1} \cdot \nabla \rho_0 + \rho_1 \nabla \cdot \mathbf{u_0}  = 0
  \end{equation}
  \begin{equation}
    \frac{\partial \mathbf{u_1}}{\partial t} + \mathbf{u_0} \cdot \nabla \mathbf{u_1} + \mathbf{u_1} \cdot \nabla \mathbf{u_0}   + \frac{1}{4\pi \rho_0} [\nabla \mathbf{b_1} \cdot \mathbf{B_0} - \mathbf{B_0} \cdot \nabla \mathbf{b_1} + \nabla \mathbf{B_0} \cdot \mathbf{b_1} - \mathbf{b_1} \cdot \nabla \mathbf{B_0} ] + \frac{\rho_1}{\rho_0}  \mathbf{u_0} \cdot \nabla \mathbf{u_0}=- \frac{1}{\rho_0} \nabla p_1
  \end{equation}
  \begin{equation}
    \frac{\partial \phi_1}{\partial t} - \mathbf{u_0} \times \mathbf{b_1} - \mathbf{u_1} \times \mathbf{B_0} - \eta \nabla^2 \phi_1 = 0
  \end{equation}
  \begin{equation}
    \frac{\partial T_1}{\partial t } + \mathbf{u_0} \cdot \nabla T_1 + (\kappa-1) (\nabla \cdot \mathbf{u_1}) T_0 +\mathbf{u_1} \cdot \nabla T_0 + (\kappa -1) (\nabla \cdot \mathbf{u_0}) T_1 = 0.
  \end{equation}
\end{subequations}
The variables with subscript ``0'' are the stationary background fields and those with subscript ``1'' are the perturbed fields which are evolved by the code.  $p_1 = \rho_0 T_1 + \rho_1 T_0$ is the perturbed pressure, $\rho, T$ are density and temperature respectively, $\mathbf{u}$ is the velocity and the adiabatic index $\kappa = 5/3$. $\phi_1$ is the perturbed magnetic flux function from which the perturbed magnetic field $\mathbf{b_1}$ is obtained
\begin{displaymath}
\mathbf{b_1} = \frac{\partial \phi_1}{\partial y} \hat{e}_x -\frac{ \partial \phi_1}{ \partial x} \hat{e}_y
\end{displaymath}

The simulation domain is a $256\times 256$ rectangular box whose $x$/$y$ axis is along/across the current sheet. As mentioned in the introduction, the Lundquist number is defined with the half-length of the box $L$:
\begin{equation}\label{eq:lundquist_number}
  S = \frac{LV_{Au}}{\eta}
\end{equation}
In our simulations, the length of the box is fixed to be $2$ ($-1 \le x \le 1$) which means all the lengths are normalized to the half length of the current sheet. As we are studying the Sweet-Parker type current sheets, the half width of the current sheet scales with $S$ as:
\begin{equation}\label{eq:half_width}
  \frac{a}{L} =  S^{-\frac{1}{2}}
\end{equation}
so $a$ changes with the Lundquist number in our simulations. As a result, the width of the simulation domain is also changing according to the Lundquist number in order to maintain proper range and resolution in $y$ direction ($\max( |y|) \sim 10 a$). Derivatives are calculated using the 6th order Compact Finite Difference (CFD) scheme \citep{lele1992}. 
At the boundaries, projected characteristics are applied to keep the boundaries open \citep{landi2005} (see Appendix \ref{appendix:characteristics}). We use a 4th order Runge-Kutta method to advance in time.

\begin{figure}
  \centering
  \includegraphics[width=15cm,keepaspectratio=true]{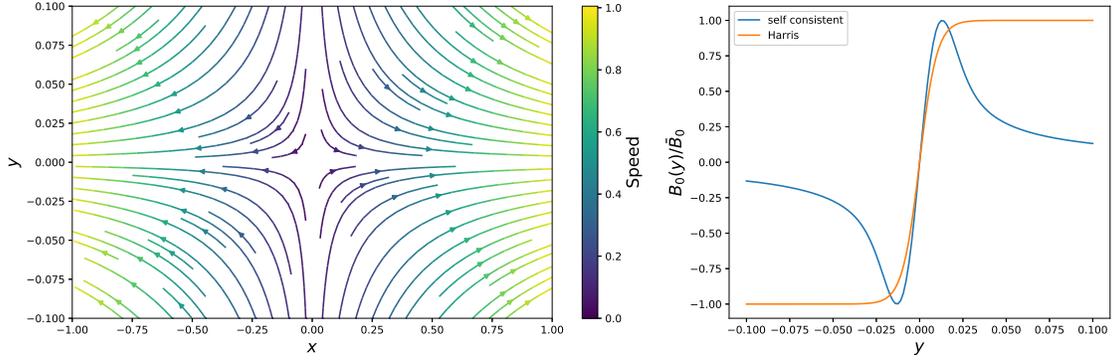}
  \caption{The background fields in the simulations with $S=10^4$. Left: Streamlines of the background flow $\mathbf{u_0}$. Right: Two types of $B_0(y)$ used in the simulations.}
  \label{fig:background_flow}
\end{figure}

The background fields are given by:
\begin{subequations}\label{backgorund_fileds}
  \begin{equation}\label{B0}
    \mathbf{B_0} = B_0(y) \hat{e}_x
  \end{equation}
  \begin{equation}\label{u0}
    \mathbf{u_0} = \Gamma x \hat{e}_x - \Gamma y \hat{e}_y
  \end{equation}
\end{subequations}
where $\Gamma$ is a constant controlling how fast the outflow is accelerated. For Sweet-Parker type current sheets, the outflow speed is the upstream Alfv\'en speed so we have
\begin{equation}\label{eq:Gamma}
  \Gamma  = \frac{V_{Au}}{L} = \tau_A ^{-1}
\end{equation}
where $\tau_A$ is the Alfv\'en crossing time of the current sheet. The plot of $\mathbf{u_0}$ is shown in the left panel of Fig (\ref{fig:background_flow}). $\rho_0$ is uniform and set as $1/4\pi$ so the local Alfv\'en speed is simply
\begin{equation}\label{eq:Alfven_speed}
  V_A(y) = B_0(y)
\end{equation}
$T_0(x,y)$ is calculated by
\begin{equation}\label{eq:T0}
  T_0(x,y) = \bar{T}_0 - \frac{1}{2} (u_{0x}^2 + u_{0y}^2 + B_0^2)
\end{equation}
so that the 0th order momentum equation is satisfied. $\bar{T}_0$ is a constant and set to be $2$ in the simulations. 
Note that self-consistency requires that $\nabla \times (\mathbf{u_0}\cdot \nabla \mathbf{u_0} - \frac{1}{4\pi \rho_0} \mathbf{B_0} \cdot \nabla \mathbf{B_0}) = 0$, 
which is satisfied by Eq (\ref{backgorund_fileds}).

The self-consistent equilibrium magnetic field $B_0(y)$ is
\begin{equation}\label{eq_dawsn}
  B_0(y) = \frac{\bar{B}_0}{0.54} \exp[-(\frac{y}{\sqrt{2}a})^2] \int_{0}^{y/\sqrt{2}a} e^{s^2}ds
\end{equation}
shown as the blue curve in the right panel of Fig (\ref{fig:background_flow}). For this equilibrium, the amplitude of $B_0$ has a maximum and then decreases toward $0$ far from the center of the sheet. In other words, because of the background inflow, the magnetic field is concentrated near the midplane of the current sheet and we use the peak value $\bar{B}_0$ (set as $1$) as the characteristic (``upstream'') magnetic field in this case. In Section (\ref{sec:num_result_dawson}) -- (\ref{sec:initial_pert}), we adopt this equilibrium field for the simulations and the linear stability analysis. In Section (\ref{sec:num_result_tanh}), we also show simulation results based on the widely-used Harris current sheet field
\begin{equation}\label{eq:harris_B0}
  B_0(y) = \bar{B}_0 \tanh (\frac{y}{a})
\end{equation}
shown as the orange curve in the right panel of Fig (\ref{fig:background_flow}) where $\bar{B}_0$ (set as $1$) is the asymptotic value of $B_0$.

Note that both $L$ and $V_{Au}$ are fixed to be $1$ thus the Lundquist number is determined solely by the resistivity $\eta$ in the code. Except for the runs shown in Section (\ref{sec:initial_pert}), we initiate the simulations with white noise, i.e. perturbations of the magnetic flux $\phi_1$ uniformly distributed in $k$ space with random phases peaking near the midplane of the current sheet.

\section{Simulation Results}\label{sec:num_result_dawson}
\begin{figure}
  \centering
  \includegraphics[width=15cm,keepaspectratio=true]{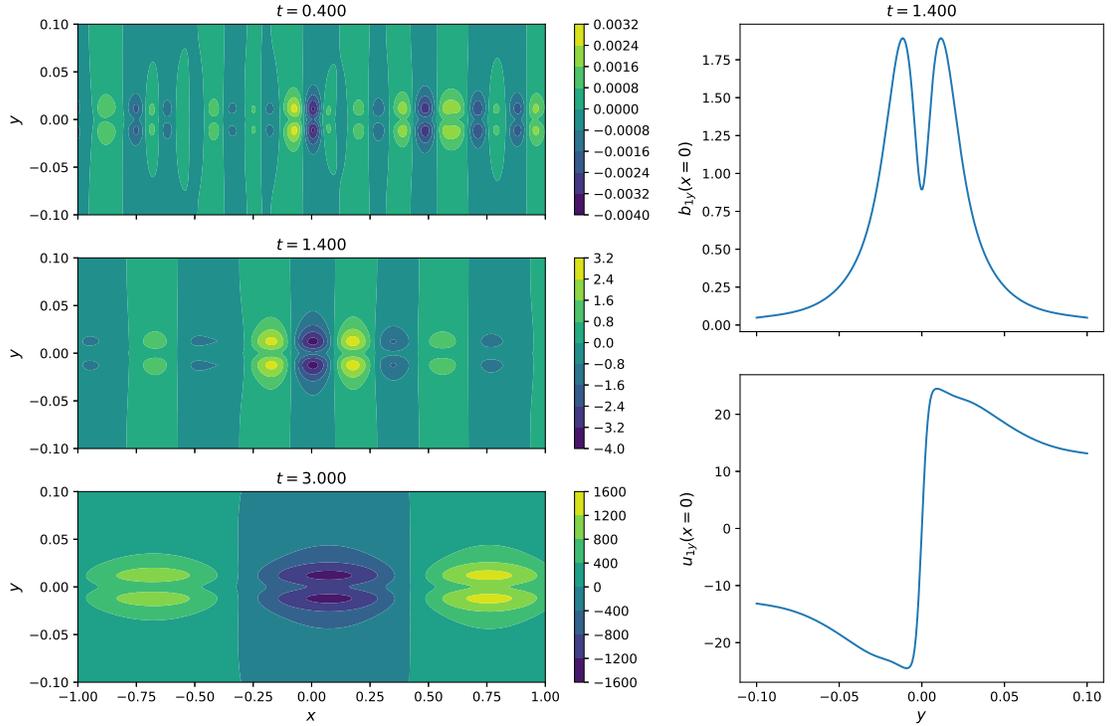}
  \caption{Left column: Snapshots of the perturbed magnetic flux function $\phi_1$ in the $S=10^4$ run. Right column: Profiles of $b_{1y}$ and $u_{1y}$ along $y$-axis at $t=1.4$.}
  \label{fig:phi_t}
\end{figure}
In this section we describe simulation results of the runs with the self-consistent background fields. Fig (\ref{fig:phi_t}) left column shows the evolution of the perturbed magnetic flux function $\phi_1$ in the run $S=10^4$ ($a=0.01$). From the top to bottom panels, we observe that magnetic islands form and grow in amplitude. At the same time, they are stretched along the outflow direction and the islands near the left and right boundaries of the simulation box are ejected out of the domain. The right column of Fig (\ref{fig:phi_t}) shows the profiles of $b_{1y}$ and $u_{1y}$ along the $y$-axis at $t=1.4$. As will be shown in Section (\ref{sec:linear_analysis}), these are essentially identical to the eigenfunctions which may be calculated from the linear theory by incorporating the expanding wavelength assumption
\begin{equation}\label{eq_kt}
  k(t) = k_0 \exp(-\Gamma t);
\end{equation}
This ansatz was first proposed by \citet{bulanov1978} to transform the problem into a 1D eigenvalue calculation for each value of $k$ for the tearing mode in a current sheet with a linearly accelerating outflow $u_{0x}=\Gamma x$. Note, however, that in the 2D current sheet, it is impossible to strictly define a purely monochromatic wave because of the lack of periodicity along $x$. In any case, the 1D assumption takes the stretching effect of the background flow properly into account: imagine two points separated by one wavelength $\lambda(t)$ along the flow
\begin{equation}
x_2(t) = x_1(t) + \lambda(t)
\end{equation}
After time $d t$, the distance between the two points becomes
\begin{equation}
\begin{aligned}
  \lambda(t+dt)  & = x_2 - x_1  + [u_{0x}(x_2) - u_{0x}(x_1)] dt \\
    & = \lambda(t) + \frac{d u_{0x}}{dx} \lambda(t)  dt
\end{aligned}
\end{equation}
which means
\begin{equation}
\frac{1}{\lambda}  \frac{d\lambda}{dt} = \frac{d u_{0x}}{dx}
\end{equation}
and we easily get Eq (\ref{eq_kt}). The result shown in Fig (\ref{fig:phi_t}) is qualitatively consistent with the time-dependent wave number assumption: we will discuss this in more detail in Section (\ref{sec:linear_analysis}).

\begin{figure}
  \centering
  \includegraphics[width=15cm,keepaspectratio=true]{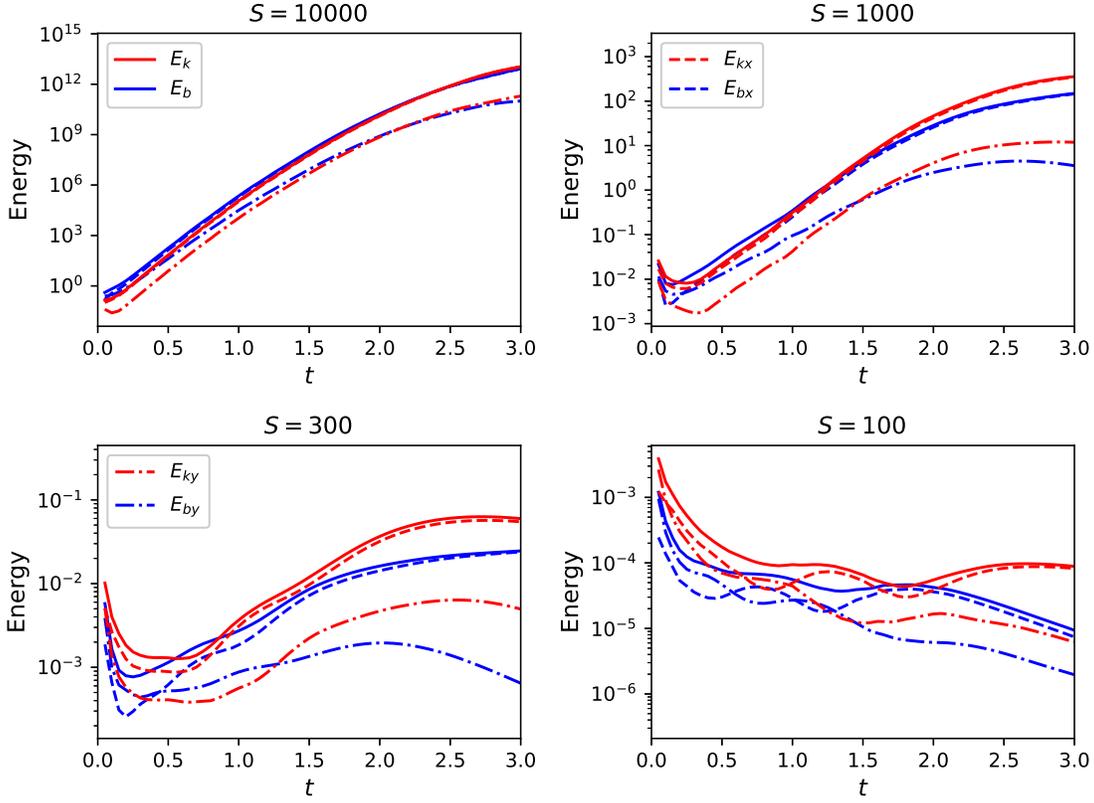}
  \caption{Energy of the perturbed velocity and magnetic fields as functions of time for different $S$. Blue lines are magnetic energy and red lines are kinetic energy. Dashed lines are the energy in $x$ components, dashed-dotted lines are the energy in $y$ components and solid lines are the total energy.}
  \label{fig:energy_time}
\end{figure}

We calculate the perturbed magnetic and kinetic energies inside the simulation domain
\begin{equation}\label{eq_Ek_Eb}
  E_{k} = \int \frac{1}{2} (u_{1x}^2 + u_{1y}^2 ), \quad E_b = \int \frac{1}{2} (b_{1x}^2 + b_{1y}^2)
\end{equation}
as functions of time. These are shown in Fig (\ref{fig:energy_time}):  from top-left to bottom-right the panels display runs for $S=10^4, 10^3, 300$ and $100$ respectively. One immediate result from Fig (\ref{fig:energy_time}) is that the growth rate of the energy in the perturbed fields decreases with decreasing Lundquist number. At $S \sim 100$, the perturbed energy becomes decreasing with time. Another feature observed in Fig (\ref{fig:energy_time}) is that the energy in the  $x$-component  of the fields dominates over the $y$-component energy. This is very clear for runs $S=10^4, 10^3$ and $300$ where the ratios $E_{kx}/E_{ky}$ and $E_{bx}/E_{by}$ tend to increase with time. This phenomenon can also be explained qualitatively by the stretching effect of the inhomogeneous outflow. Consider the 1st order magnetic field $\mathbf{b_1}$, which must be divergence free:
\begin{equation}\label{eq_divb1}
  \frac{\partial b_{1x}}{\partial x} + \frac{\partial b_{1y}}{\partial y} = 0
\end{equation}
If the perturbations are approximated by a mode with a time-dependent wave number (Eq (\ref{eq_kt})), we can estimate
\begin{equation}
  \frac{\partial b_{1x}}{\partial x} \sim k_0 b_{1x} \exp(-\Gamma t)
\end{equation}
and
\begin{equation}
  \frac{\partial b_{1y}}{\partial y} \sim \frac{b_{1y}}{\delta }
\end{equation}
where $\delta$ is the thickness of the inner layer of the tearing mode \citep{loureiro2007} and is approximately constant with time. We then get
\begin{equation}
  \frac{b_{1y}}{b_{1x}}\sim   (k_0 \delta) \exp(-\Gamma t)
\end{equation}
which means that the growth of $b_{1y}$ is slower than that of $b_{1x}$. We can apply the same analysis to $\mathbf{u_1}$ assuming that the incompressible modes dominate.

\begin{figure}
  \centering
  \includegraphics[width=10cm,keepaspectratio=true]{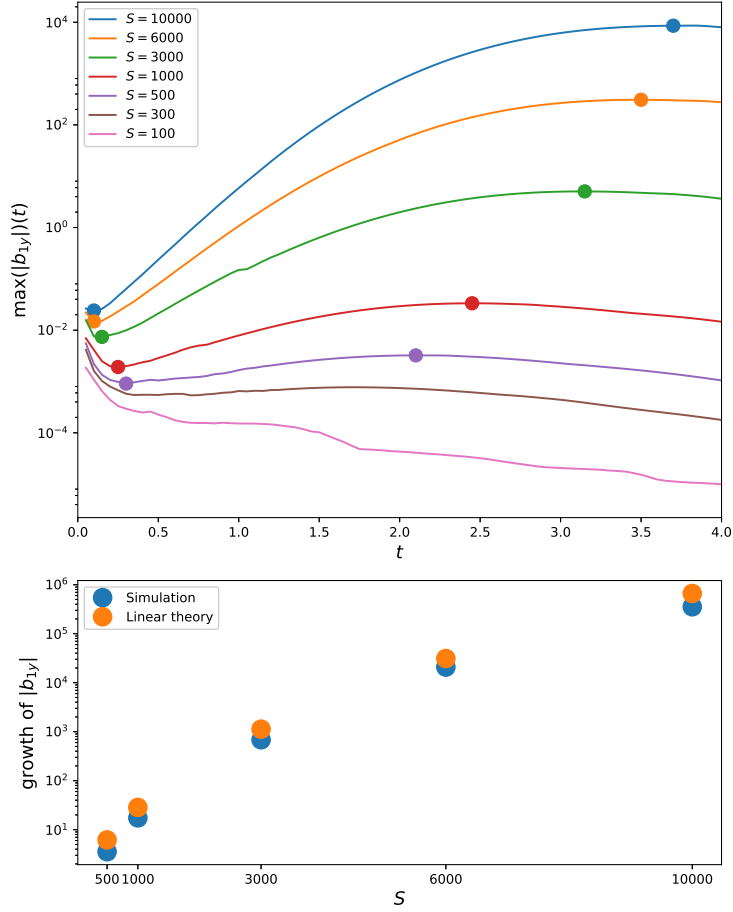}
  \caption{Upper panel: Maximum amplitude of $b_{1y}$ in the simulation domain as functions of time for several runs. The two dots on each curve mark the start and the end of the growth phase. Lower panel: Blue dots: growth of $|b_{1y}|$ during the growth phase of the simulations (the ratio between $\max (|b_{1y}|)$ values at the two dots on each curve in the upper panel) as a function of $S$. Orange dots: growth of $|b_{1y}|$ predicted by the linear theory. See $G(S)$ (Eq (\ref{eq:total_growth})) and the discussion in Section (\ref{sec:linear_analysis}).}
  \label{fig:max_by_time}
\end{figure}

In practice,  $b_{1y}$, as the reconnecting component of the field, is the more interesting quantity. Fig (\ref{fig:max_by_time}) shows the growth of $b_{1y}$ in several runs. The upper panel shows the amplitude of $b_{1y}$ as function of time for runs with different $S$. Here we use the maximum of $|b_{1y}|$ inside the whole simulation domain to represent the amplitude of $b_{1y}$. For $S < 300$, $|b_{1y}|$ does not grow with time. For the runs with $S > 300$, $|b_{1y}|$ grows after an initial transient phase and then the growth gradually slows down until reaching a maximum. The two dots on each curve mark the start and the end of the growth phase. The saturation time corresponds roughly to the time when most of the magnetic islands are ejected out. We use the ratio between the amplitudes of $b_{1y}$ at the two points to measure the total growth of $|b_{1y}|$ during a simulation, which is shown as the blue dots in the lower panel of Fig (\ref{fig:max_by_time}) as a function of Lundquist number. The orange dots in the same panel are the total growth of $b_{1y}$ estimated through the linear theory which will be discussed in next section. At $S=1000$, $|b_{1y}|$ grows by a factor of $17$ while at $S=10^4$ it grows by a factor of $\sim 3.5\times 10^5$. If we arbitrarily choose the criterion that $|b_{1y}|$ grows by a factor of $10^2$ in order to perturb the current sheet strongly, the critical Lundquist number needs to be $\sim 2000$. Here ``strongly perturb the current sheet'' means that the amplitude of the perturbation becomes large enough to trigger a fast nonlinear evolution \citep{tenerani2015,tenerani2016}.

\section{Linear stability analysis of the 2D current sheets}\label{sec:linear_analysis}
\begin{figure}
  \centering
  \includegraphics[width=15cm,keepaspectratio=true]{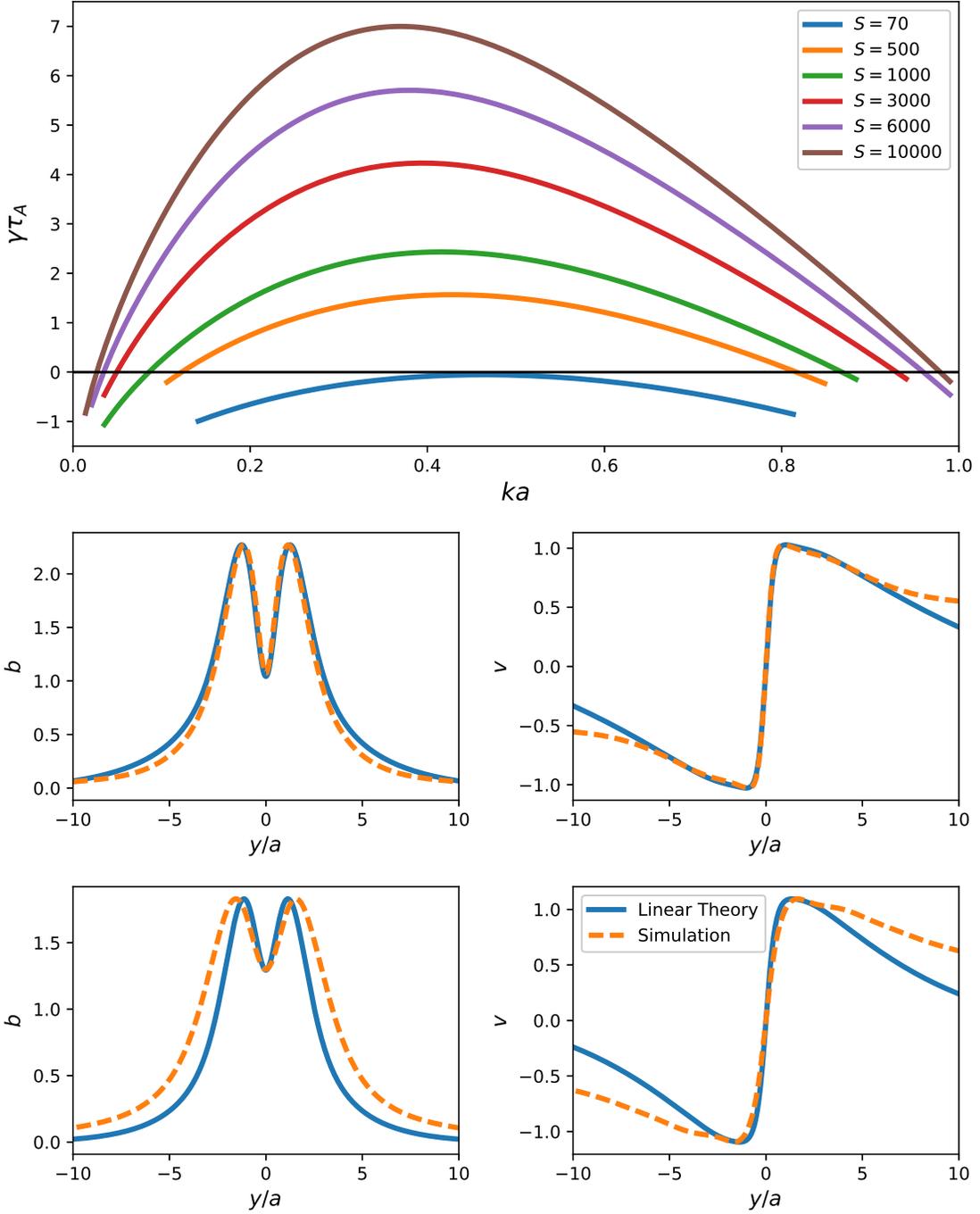}
  \caption{Linear stability analysis based on Eq (\ref{eq:v_b_BVP}). Upper panel: Dispersion relation $\gamma-k$ for different $S$. Middle panel: Blue solid line: eigenfunctions calculated through linear analysis for $S=10^4$ and $ka=0.15$. Orange dashed line: profiles of $u_{1y}$ and $b_{1y}$ taken from the simulation when $ka\approx 0.15$ (right column of Fig (\ref{fig:phi_t}), $t=1.4$). The amplitudes of the linear result and the simulation result are adjusted to be the same.  Lower panel: same as middle panel but $S=500$ and $ka=0.20$ ($t\approx 1.4$ in the simulation).}
  \label{fig:linear_theory}
\end{figure}

If we use the same background flow as used in the simulations (Eq (\ref{u0})) and include the expanding effect of the outflow on wavelengths by using a time-dependent wave vector as given by Eq (\ref{eq_kt}), assuming 
that $v$ and $b$ are functions of $y$ only (rigorously they should be functions of both $y$ and $t$), of the form 
\begin{equation}\label{eq:form_sol}
  \left(
  \begin{array}{c}
    u_{1y} \\
    b_{1y}
  \end{array}\right)
   = \left(
   \begin{array}{c}
   -i v(y)\\
   b(y)
   \end{array}\right)
    \exp \big[ i k(t) x + \int_{0}^{t} \gamma (k(t^\prime)) dt^\prime \big]
\end{equation}
we can reduce Eq (\ref{linear_MHD_eq_set}) to a set of ordinary differential equations
\begin{subequations}
  \label{eq:v_b_BVP}
  \begin{equation}
  \label{eq_v}
    y(v^{\prime\prime\prime} - k^2 v^\prime) - (\gamma + 1 ) v^{\prime\prime} + (\gamma - 1 ) k^2 v = k\frac{L}{a}[B_0(y)(b^{\prime\prime}-k^2 b)-B_0^{\prime\prime}(y)b]
  \end{equation}
  \begin{equation}
  \label{eq_b}
   (\gamma + 1)b  - yb^\prime = k  \frac{L}{a} B_0(y) v + \frac{1}{S} \, \frac{L^2}{a^2}(b^{\prime\prime}-k^2b)
  \end{equation}
\end{subequations}
In deriving Eq (\ref{eq:v_b_BVP}) we have also assumed incompressibility and adopted the following normalizations: length to the half thickness of the current sheet $a$, magnetic field to the upstream magnetic field $\bar{B}_0$, speed to the upstream Alfv\'en speed $V_{Au}$, time to Alfv\'en time $\tau_A = L/V_{Au}$ where $L$ is the half length of the current sheet. $S$ is defined as $LV_{Au}/\eta$ and we study Sweet-Parker type current sheets so that $a/L=S^{-1/2}$. Eq (\ref{eq:v_b_BVP}) is a 1D eigenvalue problem with the boundary conditions that $v$ and $b$ vanish at infinity. We numerically solve this eigenvalue problem (see Appendix \ref{appendix:series_solution}) and present the result in Fig (\ref{fig:linear_theory}). The upper panel of Fig (\ref{fig:linear_theory}) is the dispersion relation $\gamma (k)$. The shape of the $\gamma (k)$ curves is similar to that of the dispersion relation of classic tearing mode: there is a fastest growing mode $k_{f}$ for each Lundquist number and the growth rate decreases for larger or smaller $k$. Unlike the tearing mode in static current sheets whose growth rate is never negative, now part of the $\gamma(k)$ curve goes below the $\gamma =0$ line. For $S \lesssim 70$ the whole curve is below $\gamma = 0$ which means that in this configuration of the background fields, the tearing mode does not grow at all for $S \lesssim 70$.

The stabilizing effect of the background flow can be understood by looking at Eq (\ref{eq:v_b_BVP}). In Eq (\ref{eq:v_b_BVP}), the most important term that determines $\gamma$ is $(\gamma+1)v^{\prime\prime}$. In the same equation for the tearing mode inside a static current sheet, the term becomes $\gamma v^{\prime\prime}$. That is to say
\begin{equation}\label{eq:gamma_gamma_s}
  \gamma \sim \gamma_s - 1
\end{equation}
where $\gamma$ is the growth rate with the presence of flow and $\gamma_s$ is the growth rate without the flow. The difference 1($\Gamma$) comes from the time-derivative of $k$ in deriving Eq (\ref{eq:v_b_BVP}). This means that in the current sheet with the background flow, stretching of the magnetic islands due to the inhomogeneous outflow slows down their growth. Compared with the equations derived by \citep{bulanov1978} (who takes into account only the outflow), Eq (\ref{eq:v_b_BVP}) displays two major differences: (1) the coefficient in front of $v^{\prime \prime}$ changes from $\gamma + 2$ to $\gamma + 1$. (2) $y$-convection terms appear ($y(v^{\prime\prime\prime}-k^2 v^\prime)$ and $yb^\prime$). In Bulanov's equation, the constant 2($\Gamma$) consists of one $\Gamma$ which comes from $dk/dt$ (expansion of the wavelength) and another one $\Gamma$ coming from $du_{0x}/dx$ (expansion of the fluid volume) so their model is even more stable than Eq (\ref{eq:v_b_BVP}). However, when the inflow is included, the $du_{0x}/dx$ is cancelled out by $du_{0y}/dy$ because of  incompressibility. In fact, the background configuration considered by \citep{bulanov1978} is not self-consistent because their $\mathbf{u_0}=\Gamma x \hat{e}_x$ is compressible while they still assumed a uniform density. The newly-included $y$-convection terms, especially the term $y(v^{\prime\prime\prime} - k^2 v^\prime)$, though not important at large $S$, affects the calculated $\gamma$ at low $S$ ($\lesssim 10^3$) when $y(v^{\prime\prime\prime} - k^2 v^\prime)$ becomes comparable to $(\gamma + 1)v^{\prime\prime}$. We also solved Eq (\ref{eq:v_b_BVP}) without the two $y$-convection terms (not shown here) and it turns out that the growth rate becomes larger, i.e., the $y$-convection terms somewhat contribute to the stabilization of the current sheet. But as will be discussed further below, at very low $S$ the linear analysis method becomes invalid so we are unable to evaluate exactly how much stabilization the $y$-convection brings.

In the middle panel of Fig (\ref{fig:linear_theory}) we compare the eigenfunctions calculated through Eq (\ref{eq:v_b_BVP}) with the profiles taken from the simulation for $S=10^4$. The blue lines are the calculated eigenfunctions at $ka=0.15$, i.e., the wavelength $\lambda \approx 0.4$ and the orange lines are $u_{1y}$ and $b_{1y}$ along the $y$ axis taken from the $S=10^4$ run at $t=1.4$ (right column of Fig (\ref{fig:phi_t})) when the wavelength of the dominant mode is $\approx 0.4$. The amplitudes of the eigenfunctions and the cuts from the simulation are adjusted to be the same. We can see that the shapes of the eigenfunctions by the linear theory are very close to the profiles taken from the linear simulation, confirming the validity of the linear theory in this case.

\begin{figure}
  \centering
  \includegraphics[width=15cm,keepaspectratio=true]{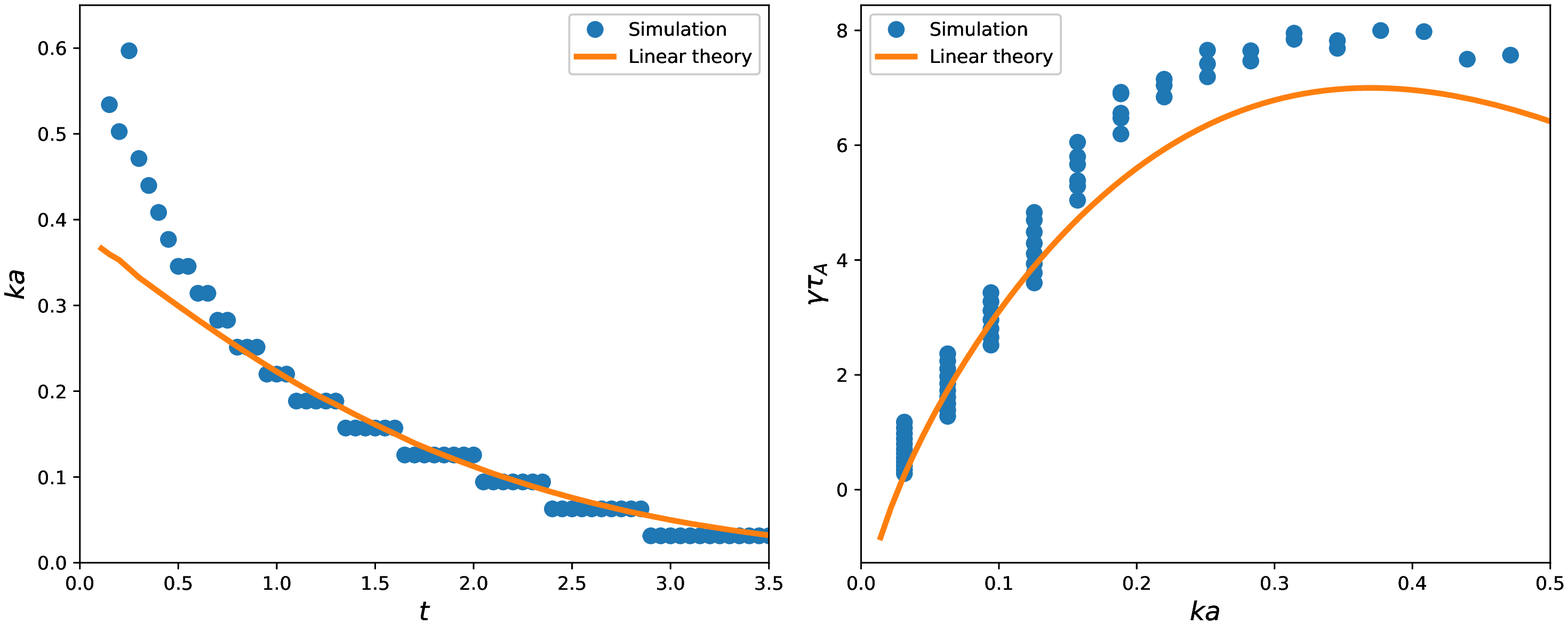}
  \caption{Left panel: Dominant wave number as a function of time. Blue dots are calculated from the simulation $S=10^4$ and the orange curve is the estimate through the linear theory. The steps in the blue dots are due to the finite resolution in $k$ when we apply Fourier transform. The orange curve is shifted in time by $\Delta t = 0.1$ because the simulation has an initial transient phase whose duration is $\sim 0.1$. Right panel: Dispersion relation $\gamma-k$. Blue dots are calculated from the simulation and the orange curve is the linear theory.}
  \label{fig:kd_t}
\end{figure}

Based on the linear theory, we are able to explain the saturation phenomenon observed in Fig (\ref{fig:max_by_time}). For a mode with wave number $k(t)$ at time $t$, the instantaneous growth rate is $\gamma(k(t))$. As time increases, $k(t)$ decreases and thus the instantaneous growth rate moves along the $\gamma-k$ curve toward left. After $k(t)$ passes the fastest-growing wave number $k_f$, the growth rate starts to decrease and eventually goes to $0$ (even below $0$). To verify this theory, we first estimate the dominant mode, i.e. the mode whose amplitude is the largest, at a certain time and compare it with the simulation result. We assume that the initial perturbations are uniformly distributed in $k$-space. At time $t$, the wave number $k = k(t)$ corresponds to the initial wave number $k_0 = k(t) e^t$ (in our normalization $\Gamma = 1$). At any time $0 \le t^\prime \le t$ this mode corresponds to
\begin{equation}
 k^\prime = k_0 e^{-t^\prime} = k(t) e^{t-t^\prime}
\end{equation}
and the instantaneous growth rate at $t^\prime$ is
\begin{equation}
  \gamma^\prime = \gamma(k^\prime) = \gamma[k(t)e^{t-t^\prime}]
\end{equation}
In order to get the amplitude of the mode $k(t)$, we integrate along the $\gamma - k$ curve from $k_0$ to $k(t)$:
\begin{equation}\label{eq:theory_kd}
  |b_{1y}|(t,k(t)) =  \exp\big[ \int_{0}^{t}\gamma(k \, e^{t-t^\prime}) dt^\prime \big]
\end{equation}
and then we are able to find the wavenumber $k_d(t)$ whose amplitude is the largest. In the left panel of Fig (\ref{fig:kd_t}), the orange curve shows the estimated dominant wave number through Eq (\ref{eq:theory_kd}) for $S=10^4$ and the blue dots show the dominant wave number calculated from the simulation $S=10^4$. In analyzing the simulation result, we first cut $b_{1y}(x,y,t)$ along $x$ axis, i.e. the midplane of the current sheet, and then multiply it by a Tukey window to remove the effect of non-periodicity along $x$ before we apply Fourier transform to it. The steps in the blue dots are due to the resolution in $k$ space when we do Fourier transform. We shift the orange curve by $\Delta t = 0.1$ because of the initial transient phase in the simulation. We see that the curve and the dots almost overlap for $t \gtrsim 1.0$. The large difference between them for $t \lesssim 0.5$ may be a result of the initial transient phase in the simulation. As we have the amplitude of $b_{1y}$ as a function of time from the simulation, we are able to calculate the instantaneous growth rate at a given time by
\begin{equation}\label{eq:instant_gamma}
  \gamma(t) = \frac{1}{|b_{1y}|} \frac{d|b_{1y}|}{dt}
\end{equation}
and then relate it with the calculated dominant wave number at the same time to get a $\gamma-k$ curve, which is shown in the right panel of Fig (\ref{fig:kd_t}). Again the blue dots are the simulation result and the orange curve is the dispersion relation given by the linear theory (the curve taken from Fig (\ref{fig:linear_theory})). The theory and the simulations agree with each other very well, verifying the physical picture that $\gamma$ moves along the $\gamma-k$ curve.

We estimate the total growth of $|b_{1y}|$ using the dispersion relation shown in Fig (\ref{fig:linear_theory}) by doing the integral along the $\gamma -k$ curves:
\begin{equation}\label{eq:total_growth}
  G(S) = \exp \bigg[ \int_{0}^{t_c}  \gamma(k_0 e^{-t^\prime}|S) dt^\prime  \bigg]
\end{equation}
where $t=0$ is at the right zero point of the $\gamma -k $ curve and $t_c$ is when $k_0 \exp(-t_c) = \pi /L$, i.e., when the wavelength reaches the length of the current sheet (generally very close to the left zero point of the curve). Note that Eq (\ref{eq:total_growth}) is actually an estimate of the upper limit of the total growth of $|b_{1y}|$ because we integrate through the whole positive part of each $\gamma-k$ curve, thus the estimate is larger than the real growth. The calculated $G(S)$ is shown as the orange dots in the lower panel of Fig (\ref{fig:max_by_time}). We can see that it is in accordance with the simulation though a little bit larger and in general the theory result and the simulation result differ by a factor of smaller than $2$.

Although the linear stability analysis based on the $k=k_0 \exp (-\Gamma t)$ assumption works quite well for $S \gtrsim 10^3$, we must point out that, for small Lundquist numbers, this assumption becomes invalid because of the small growth rate ($\gamma \lesssim \tau_{A}^{-1}$), i.e., when the growth time scale for a magnetic island becomes larger than the time scale for it to be ejected out of the current sheet. Actually, when $\gamma < \Gamma(=\tau_A^{-1})$, we cannot assume that the functions $v$ and $b$ in Eq (\ref{eq:v_b_BVP}) are independent of time anymore. Note that in Eq (\ref{eq:v_b_BVP}) $k$ is a function of time so mathematically the assumption Eq (\ref{eq:form_sol}) is not rigorous. Stated otherwise, we are unable to pose a 1D eigenvalue problem for a 2D Sweet-Parker type current sheet at very small Lundquist numbers. In the lower panel of Fig (\ref{fig:linear_theory}), we compare the simulation result with the eigenfunctions for $S=500$ and $ka=0.20$. Clearly the simulation result deviates from the eigenfunctions. There is no doubt that as $S$ becomes smaller the difference between the simulation and the approximate 1D linear theory will become larger. The linear theory predicts a threshold $S_c \approx 70$ below which $b_{1y}$ does not grow while from the upper panel of Fig (\ref{fig:max_by_time}) we see that the threshold is at least $\approx 300$.

\section{The effect of the initial perturbation}\label{sec:initial_pert}
In this section we discuss the stability of the current sheet in terms of initial conditions, i.e. in terms of the  initial perturbation. Our simulation is linear so that the amplitude of the initial noise is arbitrary. In Section (\ref{sec:num_result_dawson}) we chose the criterion for an unstable current sheet to be that $|b_{1y}|$ grows by a factor of $10^2$ which gives a critical Lundquist number $\sim 2000$. However, that is not a universal criterion, because depending on the initial state,  we must allow $|b_{1y}|$ to grow
sufficiently that the sheet is strongly affected, i.e. nonlinear evolution is triggered, at times comparable to their evacuation from the domain.

\begin{figure}
  \centering
  \includegraphics[width=15cm,keepaspectratio=true]{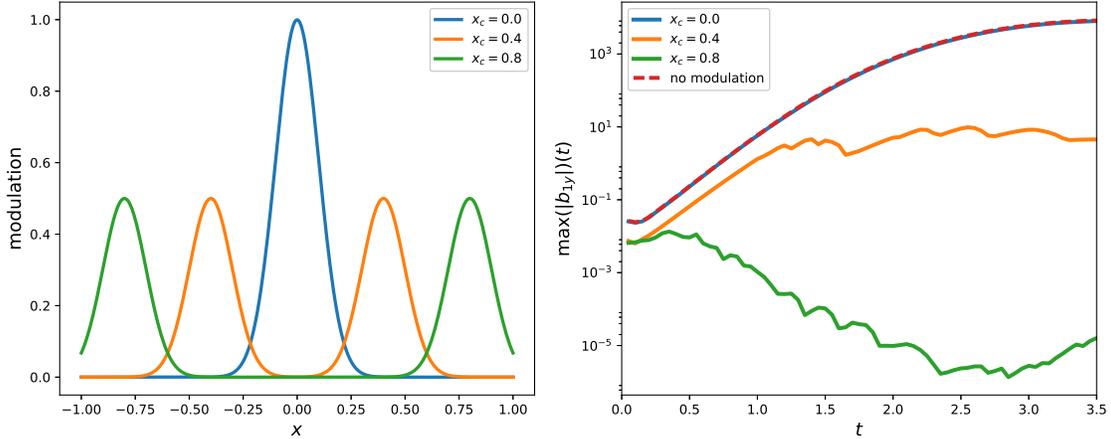}
  \caption{Left: Modulation functions multiplied to the initial perturbations in the three runs with the identical configurations as the run $S=10^4$ shown in Fig (\ref{fig:phi_t}). Right: Amplitudes of $b_{1y}$ as functions of time in the three runs with the modulation functions shown in the left panel. The red dashed line is the original run without modulation, i.e. the $S=10^4$ curve in the upper panel of Fig (\ref{fig:max_by_time}).}
  \label{fig:max_by_time_diff_init}
\end{figure}

Because the configuration of the current sheet in this study is fully 2D and the simulation is open-boundary, it is natural to propose that the shape of the initial perturbation also affects the stability. To be more specific, where the initial perturbation is excited on the current sheet is important because the generation location of a magnetic island determines how long the island grows before it is ejected out of the current sheet. Here the location refers to the distance to $x=0$, i.e. the location along the outflow. Based on the $S=10^4$ run, we carried out three more runs with identical background configuration but for each run we modulate the initial perturbation by multiplying it with a gaussian function:
\begin{equation}\label{eq:modulation_functions}
  f(x|x_c) = \frac{1}{2} \big[ \exp(- \frac{(x-x_c)^2}{2\sigma^2}) +\exp(- \frac{(x+x_c)^2}{2\sigma^2})  \big]
\end{equation}
where $x_c = 0.0$, $0.4$ or $0.8$ and $\sigma = 0.1$. The 3 modulation functions are plotted in the left panel of Fig (\ref{fig:max_by_time_diff_init}). Through the modulation functions we are able to localize the initial perturbation around $\pm x_c$. We calculate the amplitudes $b_{1y}$ as functions of time for each of the three runs and present the results in the right panel of Fig (\ref{fig:max_by_time_diff_init}). The red dashed line shows the result of the original run without modulation, i.e. the $S=10^4$ curve in the upper panel of Fig (\ref{fig:max_by_time}). We see that the curve of the $x_c = 0.0 $ run almost overlaps with the original run, indicating that in the simulation where the initial perturbation is uniformly distributed along the current sheet, the dominant magnetic islands are those generated near the center of the current sheet,  as clearly seen in the left column of Fig (\ref{fig:phi_t}). In the $x_c=0.4$ run, the growth rate of $|b_{1y}|$ is almost the same as the $x_c =0.0$ run at short times ($t \lesssim 1.0$) but later the growth rapidly stops. As a result, the total growth of $|b_{1y}|$ in this run is a factor of $\sim 1.5\times 10^3$ rather than the factor $3.5\times 10^5$ of the $x_c=0.0$ run. For $x_c = 0.8$, there is nearly no growth phase for $|b_{1y}|$. These results agree with the idea that the evacuation time of the magnetic island is a key factor in determining how much the island can grow. Assuming that a magnetic island is generated at $x=x_c$, we can estimate the evacuation time of the magnetic island
\begin{equation}\label{eq:evacuation_time}
\begin{aligned}
  \tau_{e}(x_c) & = \int_{x_c}^{1} \frac{dx}{u_{0x}(x)} = \int_{x_c}^{1} \frac{dx}{x} \\
  & = \ln(\frac{1}{x_c})
\end{aligned}
\end{equation}
We then get $\tau_{e}(0.4)=0.92$ and $\tau_{e}(0.8) = 0.22$ which agree with the simulation results.

\section{Simulation results with the Harris current sheet model}\label{sec:num_result_tanh}
\begin{figure}
  \centering
  \includegraphics[width=15cm,keepaspectratio=true]{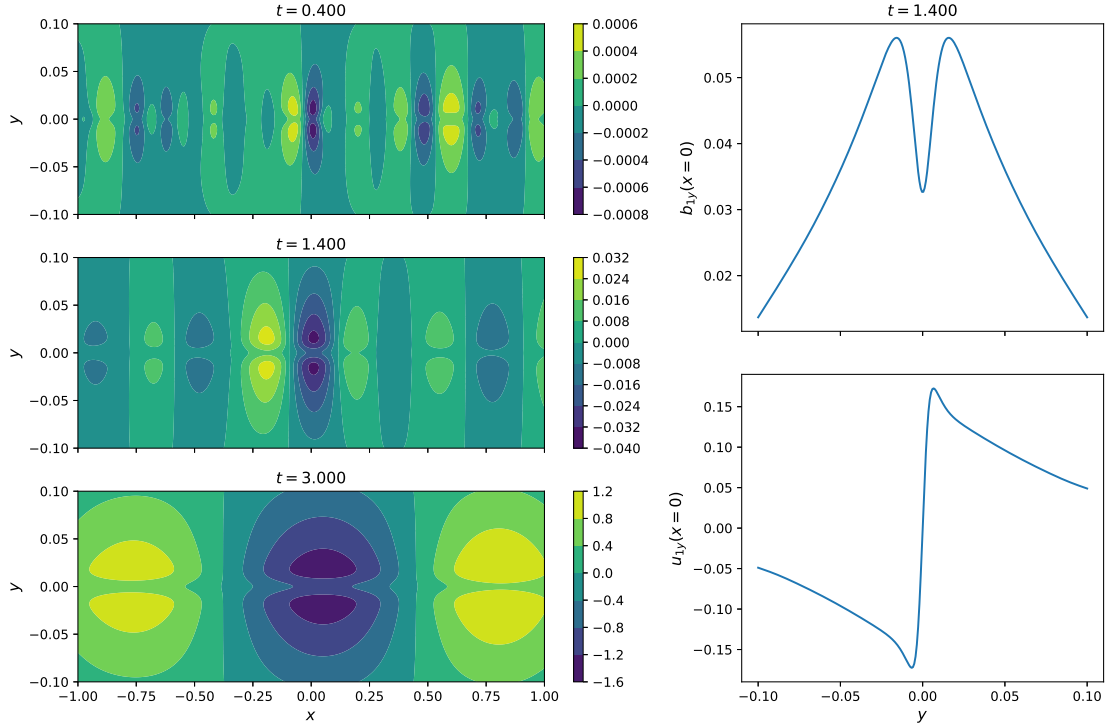}
  \caption{Left column: Snapshots of the perturbed magnetic flux function $\phi_1$ in the $S=10^4$ run with a Harris current sheet field. Right column: Profiles of $b_{1y}$ and $u_{1y}$ along $y$-axis at $t=1.4$.}
  \label{fig:phi_t_tanh}
\end{figure}

\begin{figure}
  \centering
  \includegraphics[width=10cm,keepaspectratio=true]{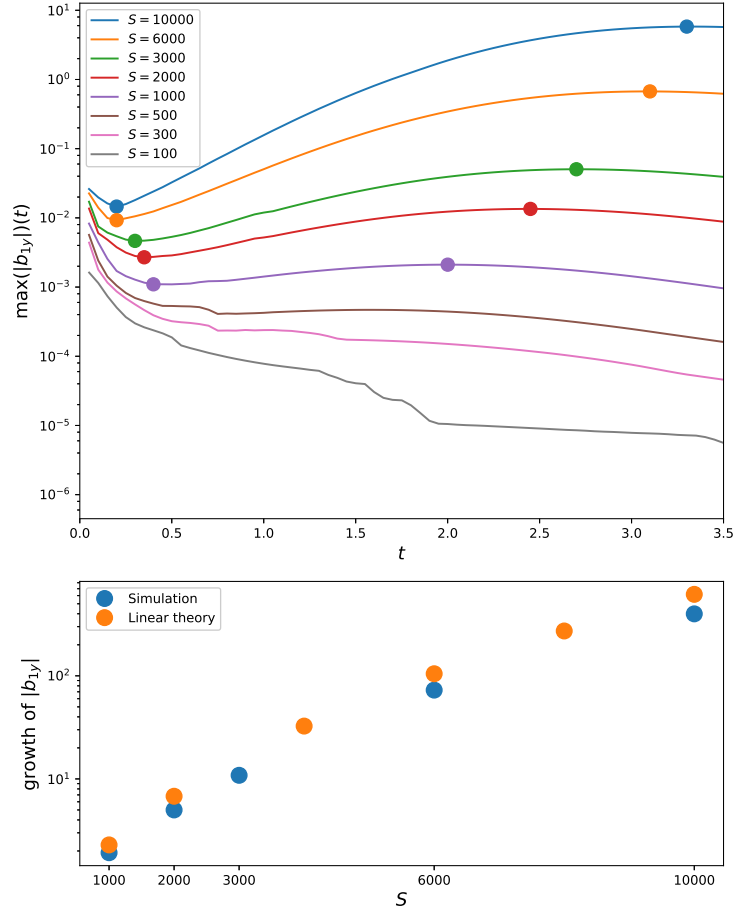}
  \caption{Same as Fig (\ref{fig:max_by_time}) except for the use of the Harris current sheet field.}
  \label{fig:max_by_time_tanh}
\end{figure}

\begin{figure}
  \centering
  \includegraphics[width=15cm,keepaspectratio=true]{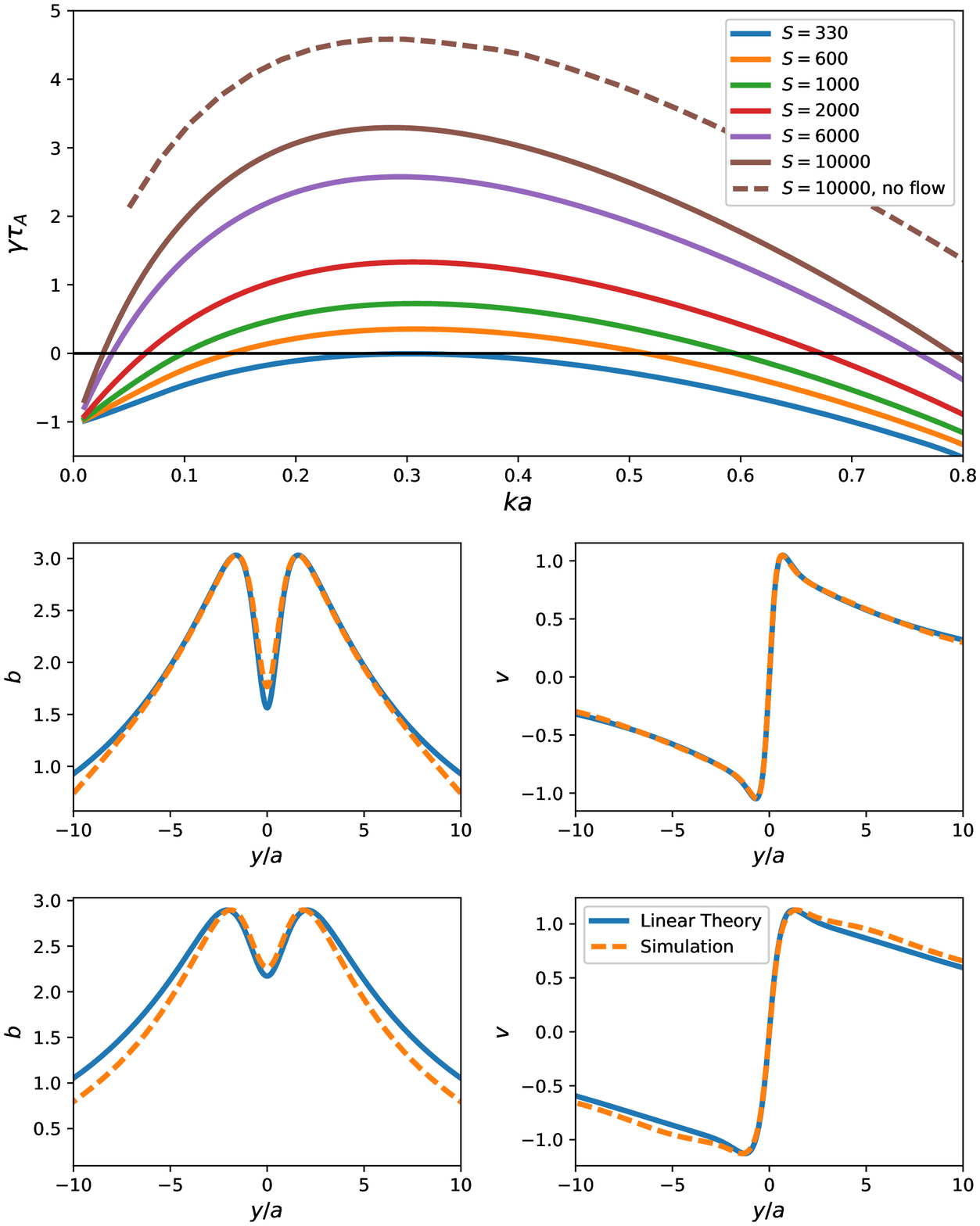}
  \caption{Linear stability analysis based on Eq (\ref{eq:v_b_BVP}) and $B_0=\tanh (y/a)$. Upper panel: Dispersion relation $\gamma-k$ for different $S$. Middle panel: Blue solid line: eigenfunctions calculated through linear analysis for $S=10^4$ and $ka=0.15$. Orange dashed line: profiles of $u_{1y}$ and $b_{1y}$ taken from the simulation when $ka\approx 0.15$ (right column of Fig (\ref{fig:phi_t_tanh}), $t=1.4$). The amplitudes of the linear result and the simulation result are adjusted to be the same.  Lower panel: same as middle panel but $S=500$ and $ka=0.20$ ($t\approx 1.4$ in the simulation).}
  \label{fig:linear_theory_tanh}
\end{figure}

We carry out the same simulations as those in Section (\ref{sec:num_result_dawson}) but change $B_0$ to the Harris current sheet field (Eq (\ref{eq:harris_B0})) and present the results in Fig (\ref{fig:phi_t_tanh}) and (\ref{fig:max_by_time_tanh}). We also carry out the simplified 1D linear stability analysis with the Harris current sheet field, shown in Fig (\ref{fig:linear_theory_tanh}) where the dashed line is the result for a static Harris current sheet and is plotted to show the stabilizing effect of the background flow (see Eq (\ref{eq:gamma_gamma_s}) and the discussion in Section (\ref{sec:linear_analysis})). 
Note that we are still assuming the background fields are stationary though technically speaking the velocity field Eq (\ref{u0}) and the Harris magnetic field Eq (\ref{eq:harris_B0}) do not satisfy the induction equation. Indeed the 0th order fields should vary with time and their evolution will hence affect the evolution of the 1st order quantities. 
In this section, however, we ignore the non-self-consistency of the background fields to illustrate the dependence of the current sheet stability on different background magnetic fields. 
In reality, this implies that our results are physically significant only for growth rates large enough ($\gamma \tau_A > 1$) where the evolution of the background fields is negeligible.

Compared with the self-consistent $B_0$ case, the growth rate with $B_0(y)=\bar{B}_0 \tanh(y/a)$ is smaller. By comparing Fig (\ref{fig:phi_t_tanh}) with Fig (\ref{fig:phi_t}), we see that the magnetic islands are much thinner with the self-consistent field because of the more concentrated $B_0$ and that is why the growth rate with the self-consistent $B_0$ is larger. The growth of $|b_{1y}|$ is also much smaller than before: at $S=1000$, $|b_{1y}|$ grows by a factor of $1.9$ ($17$ for self-consistent field) while at $S=10^4$ it grows by a factor of $400$ ($3.5\times 10^5$ for self-consistent field). If we adopt the same criterion for an unstable current sheet as in Section (\ref{sec:num_result_dawson}) that $|b_{1y}|$ grows by a factor of $10^2$, the critical Lundquist number is at least $\sim 6000$ ($2000$ for self-consistent field). Despite of the quantitative differences, the qualitative properties we discuss in Section (\ref{sec:num_result_dawson}), (\ref{sec:linear_analysis}) and (\ref{sec:initial_pert}), i.e. the expanding wavelength and the finite growth time effect, are also applicable here.

\section{Conclusion}\label{sec:conclusion}
In this study, we have explored the tearing instability inside 2D Sweet-Parker type current sheets at low Lundquist numbers using both 2D linear open-boundary MHD simulations and a simplified 1D linear stability analysis. We find that in the linear stability analysis, the assumption of a time-dependent wave number $k=k_0e^{-\Gamma t}$ proposed by \citep{bulanov1978} works quite well for $S \gtrsim 10^3$. As predicted by the linear theory, the simulations show that the inhomogeneous background outflow stretches the growing magnetic islands and the stretching slows down the island growth. On the other hand, because of the finite size of the current sheet, the magnetic islands are evacuated out of the current sheet within a finite time (several Alfv\'en times) and thus they cannot grow infinitely (as would happen in a periodic current sheet). In the simulations initialized with white noise we clearly observe a saturation phase after the growing phase, corresponding to the time when most of the magnetic islands are ejected out. As a consequence, the excitation location of the initial perturbation is an important factor in determining how much perturbations can grow.

We estimate through the simulations the critical Lundquist number above which the current sheet can be viewed as ``strongly perturbed''. The result is dependent on the shape of the background magnetic field $B_0(y)$ we use. Generally, the Harris current sheet field is more stable than the self-consistent background field (Eq (\ref{eq_dawsn})). Based on the criterion that $|b_{1y}|$ grows by a factor $10^2$ during the growth phase, the Lundquist number threshold can be $\sim 2000$ or $\sim 6000$ depending on which $B_0(y)$ is used. We may say that under this specific criterion the threshold $S_c$ is several thousands. However, no universal criterion which is independent of context can be defined, since amplitude and localization of fluctuations in the initial current are fundamental to the subsequent evolution.

We find that the flow structure of the SP sheet adequately accounts for the low $S$ stabilization without the need of other factors, such as proposed by \citep{loureiro2013}. Their suggested instability threshold, given by
\begin{equation}\label{eq:unstable_threshold_loureiro2013}
  \frac{\delta}{a}  = f
\end{equation}
where $\delta$ is the thickness of the inner layer, $a$ is the thickness of the current sheet, and $f \simeq1/3$ is a constant, yields a critical $S$ compatible with findings of some simulations.
On the other hand, the scaling of the stabilization
criterion with current sheet aspect ratio appears to go in the wrong direction.
Indeed, consider a current sheet whose aspect ratio is
\begin{equation}\label{eq:aspect_ratio}
  \frac{a}{L} \sim S^{-\alpha}.
\end{equation}
The asymptotic scaling of $\delta$ then becomes (see e.g. \citep{tenerani2016})
\begin{equation}
  \frac{\delta}{a} \sim S^{-\frac{1}{4}(1-\alpha)}
\end{equation}
and the threshold Lundquist number $S_c$ from the criterion above would become
\begin{equation}\label{eq:critical_S_loureiro2013}
  S_c = (\frac{1}{f})^{\frac{4}{1-\alpha}}
\end{equation}
For a Sweet-Parker type current sheet ($\alpha=1/2$) and $f=1/3$, Eq (\ref{eq:critical_S_loureiro2013}) gives $S_c \approx 6.5\times 10^3$, close to $10^4$. However, $S_c$ given by Eq (\ref{eq:critical_S_loureiro2013}) is an increasing function of $\alpha$, which means that the thinner a current sheet the more stable it would be, contrary to all numerical and physical intuition.
The point is that at low Lundquist number ($<10^4$) asymptotic scalings really do not work in the first place. On the other hand, we have demonstrated that velocity fields indeed provide the right criteria for stability.

In terms of astrophysical structures, such as for example coronal magnetic fields or prominence eruptions,
fluctuations are always present, propagating from other coronal regions or from the photosphere. As shown by \citet{tenerani2016}, at very high Lundquist numbers 2D current sheets tend to disrupt in a quasi self-similar way, in which case flurrys of smaller scale current sheets are generated iteratively. At each step, the Lundquist number decreases, and finally they will cross the critical $S$ at which further disruption is quenched. However, at those scales where $S$ becomes small enough, kinetic effects, explored in \citep{pucci2017}, become fundamental. We plan to address this question in subsequent works.

\acknowledgments
We would like to thank Fulvia Pucci for many useful discussions. This research was supported by the NSF-DOE Partnership in Basic Plasma Science and Engineering award n. 1619611 and the NASA Parker Solar Probe Observatory Scientist grant NNX15AF34G. This work used the Extreme Science and Engineering Discovery Environment (XSEDE) \citep{towns2014} Comet at the San Diego Supercomputer Center through allocation TG-AST160007. XSEDE is supported by National Science Foundation grant number ACI-1548562.


%\vspace{5mm}
%\facilities{HST(STIS), Swift(XRT and UVOT), AAVSO, CTIO:1.3m,


%\software{astropy \citep{2013A&A...558A..33A},
%          Cloudy \citep{2013RMxAA..49..137F},
%          SExtractor \citep{1996A&AS..117..393B}
%          }



\appendix

\section{Characteristic form of the linear MHD equations}\label{appendix:characteristics}
The linear ideal MHD equation set is a partial differential equation system and can be written in the form:
\begin{equation}\label{eq:form_linear_MHD}
\frac{\partial \mathbf{U_1}}{\partial t} + \mathbf{A_0} \frac{\partial \mathbf{U_1}}{\partial x} + \mathbf{C_0} \frac{\partial \mathbf{U_1}}{\partial y} + \mathbf{D} = 0
\end{equation}
where $\mathbf{U_1}$ is the vector of the 1st order variables:
\begin{equation}
\mathbf{U_1} = (\rho_1, u_{1x}, u_{1y}, u_{1z}, b_{1x}, b_{1y}, b_{1z}, T_1)
\end{equation}
The coefficient matrices $\mathbf{A_0}$ and $\mathbf{C_0}$ include only the 0th order quantities and the matrix $\mathbf{D} = \mathbf{F}\mathbf{U_1}$ contains other terms which do not include any derivatives of the 1st order variables. Compared with the case of the nonlinear MHD equations, the first 3 terms on the L.H.S. of Eq (\ref{eq:form_linear_MHD}) are unchanged except that the derivatives are for 1st order variables only and the coefficients are 0th order. The matrix $\mathbf{D}$ now contains more terms than the nonlinear case because all the derivatives of the 0th order quantities are included in it, e.g. $[\mathbf{u_1} \cdot \nabla \rho_0 + (\nabla \cdot \mathbf{u_0} ) \rho_1]$ in Eq (\ref{eq:linear_MHD_eq_set:rho1}). Note that $\mathbf{D}$ does not affect the projection of characteristics because it does not have any ``waves'' of $\mathbf{U_1}$.

The calculation of the characteristics is almost the same as shown in \citep{landi2005}. For example, along $x$, we first diagonalize the matrix $\mathbf{A_0}$
\begin{equation}
\mathbf{A_0} = \mathbf{S \Lambda S^{-1}}
\end{equation}
where the diagonal matrix
\begin{equation}
\Lambda = \mathrm{diag}\{ u_{0x} + f_x, u_{0x}+a_x, u_{0x}+s_x, u_{0x}, u_{0x}-s_x, u_{0x}-a_x, u_{0x}-f_x\}
\end{equation}
consists of the speeds of the 7 MHD modes (projected along the $x$ direction). $u_{0x}$, $s_x$, $a_x$, $f_x$ are the local entropy mode speed, slow magnetosonic wave speed, Alfv\'en speed and fast magnetosonic wave speed projected along $x$. Note that when we calculate the characteristics along $x$, there is no need to consider $b_{1x}$ because of the divergence-free condition. Then we are able to decompose $\mathbf{A_0} \partial \mathbf{U_1} / \partial x$ into the linear combinations of the 7 characteristics. The result is quite lengthy and is not shown here, but note that one can write down the result easily by adding subscript ``0'' to all the variables in the coefficients and subscript ``1'' to all the variables in the derivatives for the equations shown in the appendix of \citep{landi2005}.

In order to impose open-boundary condition on a certain side of the simulation box, we decompose the corresponding derivatives into the linear combination of the characteristics, e.g., decompose the $x$-derivative terms at the right/left boundaries. For each of the 7 characteristics, we decide whether it is propagating inward or outward of the domain by looking at its mode speed. If it is outgoing, we keep this characteristic; otherwise we set it as $0$. By doing this we ensure that no information is going into the simulation domain and the boundary is ``open''.

When the resistivity is non-zero, the 2nd order derivative term will cause some reflections at the boundaries. But practically, as the resistivity is quite small, the resistive term does not influence the simulation significantly at the boundaries.

\section{Series-expansion method for solving the ordinary differential equation set}\label{appendix:series_solution}
Eq (\ref{eq:v_b_BVP}) is a 5th order homogeneous boundary value problem (BVP) with a regular singular point $y=0$. The widely used methods for BVPs, e.g. the shooting method, usually involve the integrals of the ODEs over the domain. However, for ODEs with singular points such as Eq (\ref{eq:v_b_BVP}) it is hard to evaluate the integrals. Thus we adopt the method of series-expansion to solve Eq (\ref{eq:v_b_BVP}). For simplicity, we write Eq (\ref{eq:v_b_BVP}) in the form
\begin{equation}\label{eq:form_BVP}
  \mathcal{L} (v,b | \gamma, k, S) = 0
\end{equation}
where $\mathcal{L}$ is the linear operator corresponding to the ODE set with $\gamma, k, S$ (and $B_0(y)$) as parameters. Our goal is, by imposing proper boundary conditions, to determine the dispersion relation $\gamma(k)$ and the solution $v(y)$ and $b(y)$ simultaneously.

Before discussing the series expansion method, we would like to first clarify the boundary conditions. The equation is 5th order so 5 boundary conditions are needed. We set the domain to be $y \in [0,L_y]$ where $L_y = 17$ so that the right boundary is far enough to impose the asymptotic boundary conditions. We expect $v$ to be odd and $b$ to be even in $y$ so that
\begin{equation}\label{eq:BVP_bc_left}
  v(0) = 0 ,\quad b^\prime (0) = 0
\end{equation}
For large enough $y$, $B_0(y)$ (Eq (\ref{eq_dawsn})) can be approximated by
\begin{equation}\label{eq:B0_asymp}
  B_0(y) \approx 1.31  \times \frac{1}{y}   + O(1/y^3)
\end{equation}
Far from the center of the current sheet, the perturbations asymptote to $0$. Plugging Eq (\ref{eq:B0_asymp}) in Eq (\ref{eq:v_b_BVP}) and eliminating $v$, we get a 5th order ODE for $b$, which by keeping only the 5th and 4th order derivative terms a priori, becomes
\begin{equation}\label{eq:asymp_b}
  b^{(5)} +  (yb)^{(4)} = 0
\end{equation}
which gives the solution
\begin{equation}\label{eq:asmp_b_2}
  b(y) \propto  \exp (-\frac{1}{2}y^2)
\end{equation}
and we further get
\begin{equation}\label{eq:asmp_v}
  v(y) = \frac{1}{1.31} \times S^{-\frac{1}{2}}  \frac{k^2 + \gamma +2}{k} y b(y)
\end{equation}
The 3 boundary conditions at the right boundary are then given by
\begin{subequations}\label{eq:BVP_bc_right}
  \begin{equation}
    b^\prime + L_y b = 0
  \end{equation}
  \begin{equation}
    v^\prime + (L_y - \frac{1}{L_y})v = 0
  \end{equation}
  \begin{equation}
    v - \frac{1}{1.31}  S^{-\frac{1}{2}}  \frac{k^2 + \gamma +2}{k} L_y b = 0
  \end{equation}
\end{subequations}
Eq (\ref{eq:BVP_bc_left}) and (\ref{eq:BVP_bc_right}) are the 5 boundary conditions for the BVP. Note that Eq (\ref{eq:BVP_bc_right}) are not equivalent to open-boundary conditions adopted in the simulations unless $L_y \rightarrow \infty $ where all 1st order quantities vanish. But as the simulation box is large enough $\max(|y|) \sim 10 a$ and the initial perturbations are localized around $y = 0$, the difference in the boundary conditions is negligible and we are still allowed to make comparisons between the linear theory and the simulations.

We expand $v(y)$, $b(y)$ and $B_0(y)$ in power series of $y$ around a point $y_c$
\begin{equation}\label{eq:series_expansion}
  v(y) = \sum_{n=0}^{N} v_n (y-y_c)^n , \quad b(y) = \sum_{n=0}^{N} b_n (y-y_c)^n, \quad B_0(y) = \sum_{n=0}^{N} B_n (y-y_c)^n
\end{equation}
where $v_n$ and $b_n$ are quantities to be determined and $N=100$ is the cut-off order of the series. $B_n$ are given by the Taylor expansion of $B_0(y)$ around $y_c$. Once we get the values of $v_n$ and $b_n$, the solution $v(y)$ and $b(y)$ can be recovered. We plug Eq (\ref{eq:series_expansion}) into Eq (\ref{eq:v_b_BVP}) and get the recurrence relations by matching the coefficients in front of $(y-y_c)^n$:
\begin{subequations}\label{eq:recur_rel}
  \begin{equation}\label{eq:recur_rel_v}
  \begin{aligned}
    y_c (n+3)(n+2)(n+1) &v_{n+3} = (n+2)(n+1)(\gamma +1-n)v_{n+2} + k^2 y_c (n+1)v_{n+1} - (\gamma-1-n)k^2 v_n \\
                    & + k S^{\frac{1}{2}} \sum_{l=0}^{n} \big[(l+2)(l+1)B_{n-l} b_{l+2} - k^2 B_{n-l}b_l - (n-l+2)(n-l+1)B_{n-l+2}b_l \big]
  \end{aligned}
  \end{equation}

  \begin{equation}\label{eq:recur_rel_b}
    (n+2)(n+1)b_{n+2} = (\gamma+1-n+k^2)b_n - y_c(n+1)b_{n+1} - kS^{\frac{1}{2}} \sum_{l=0}^{n} B_{n-l}v_l
  \end{equation}
\end{subequations}
From Eq (\ref{eq:recur_rel}) it is obvious that there are only 5 free parameters ($v_0,v_1,v_2,b_0,b_1$): all the $v_n$ for $n\ge 3$ and $b_n$ for $n\ge 2$ can be expressed in the 5 free parameters and so for $v(y)$ and $b(y)$:
\begin{equation}\label{eq:v_b_5_params}
  v(y) = v(y|v_0,v_1,v_2,b_0,b_1), \quad b(y) = b(y|v_0,v_1,v_2,b_0,b_1)
\end{equation}
Then the 5 boundary conditions can be written as a linear algebraic equation set
\begin{equation}\label{eq:BVP_BC_5_params}
  \mathbf{M}(\gamma,k,S) \mathbf{V} = 0
\end{equation}
where $\mathbf{V}=(v_0,v_1,v_2,b_0,b_1)^T$ and $\mathbf{M}$ is a matrix which is a function of $\gamma$, $k$, $S$. In order to get a nontrivial solution, we need
\begin{equation}\label{eq:BVP_det}
  \det (\mathbf{M}) = 0
\end{equation}
which gives the dispersion relation $\gamma(k,S)$ and then we get $v(y)$ and $b(y)$.

In practice, we must do the expansion at multiple points ($y_c$) to ensure convergence of the series. If we use $N_c$ points to expand the functions, we need $5 N_c$ equations to determine the solution. Apart from the $5$ boundary conditions Eq (\ref{eq:BVP_bc_left}) and (\ref{eq:BVP_bc_right}), the other $5(N_c - 1)$ equations come from the continuities of $v(y)$, $v^\prime(y)$, $v^{\prime\prime}(y)$, $b(y)$, $b^\prime(y)$ at $N_c -1$ junction points ($y_m$), each of which lies between two neighbouring expansion points. In solving Eq (\ref{eq:v_b_BVP}) we use $11$ expansion points.

\end{CJK*}
\end{document}